\documentstyle[sprocl,epsfig]{article}

\def\be{\begin{equation}}
\def\ee{\end{equation}}
\def\bea{\begin{eqnarray}}
\def\eea{\end{eqnarray}}

\def\tbst{\vrule height2.3ex depth1ex width0pt}%
\def\beq{\begin{equation}}
\def\eeq{\end{equation}}

\def\unbr{\underbrace}

\begin{document}

\title{dE/dx PARTICLE IDENTIFICATION FOR COLLIDER DETECTORS}

\author{Hitoshi Yamamoto}

\address{Dept. of Physics and Astronomy \\
  The University of Hawaii \\
  2505 Correa Rd., Honolulu, HI 96822, USA}


\begin{flushright}
  UH-511-943-99
\end{flushright}

\vspace{0.5in}
\maketitle\abstracts{
We review some basic features of $dE/dx$ particle identification
that are relevant to high energy physics tracking devices.
Gas-based drift chambers as well as silicon trackers are
discussed. 
}
  
\section{Model of ionization}

One of the most successful models of ionization is the photo-absorption 
ionization model~\cite{AllisonCobb}. The model assumes that the process of interest
is dominated by single-photon exchange which allows one to relate the ionization
process to the photo-absorption process for which the data is more readily available.
Also the model is restricted to non-relativistic recoil electron, thus it does not
include high-energy knock-on electrons ($\delta$ rays). 
In this model, the single collision
cross section can be written as
\begin{eqnarray}
  {d\sigma\over dT} &=&
    {\alpha\over\beta^2\pi}\Bigg[
     {\sigma_\gamma\over TZ}
    \ln{2m_e\beta^2/T
      \over\sqrt{(1-\beta^2\epsilon_1)^2 + \beta^4\epsilon_2^2}} \nonumber\\
   &&\qquad\qquad\qquad
     + \unbr{
     {\Theta\over n_e}\Bigg(\beta^2-{\epsilon_1\over|\epsilon|^2}\Bigg)
        }_{\hbox{Cerenkov}}
     + \unbr{{1\over T^2}\int_0^T dT
      {\sigma_\gamma\over Z}
       }_{\hbox{Rutherford}}\; \Bigg]\,,
  \label{eq:AllisCobb}
\end{eqnarray}
where the first term results in the logarithmic rise of ionization, the second term
is the Cerenkov term, and the last term is the Rutherford scattering off
a quasi-free electron. The parameters are:
\begin{center}
  \begin{tabular}{rlrl}
     $T$ & energy loss per collision &
     $\sigma_\gamma(T)$ & photo abs. cross section \\
     $m_e$ & electron mass &
     $n_e$              & electron density \\
     $\alpha$ & $1/137$ &
     $\beta$ & velocity of projectile \\
     $Z$   & atomic number of material &
   $\epsilon_1 + i\epsilon_2$ & $\equiv\epsilon$ : dielectric constant\\
    $\Theta$ & $= \arg(1-\epsilon^*\beta^2)$
  \end{tabular}
\end{center}
The logarithmic rise can be seen in the first term
by taking the limit of $\beta\to 1$ and
low density $\epsilon \equiv \epsilon_1 + i\epsilon_2\to 1$:
\beq
  \ln{1\over\sqrt{(1-\beta^2\epsilon_1)^2 + \beta^4\epsilon_2^2}} \to
  \ln{1\over 1-\beta^2} = \ln \gamma^2\,.
 \label{eq:lowden}
\eeq

On the other hand, at large photon energy $T$ (far above
poles), the dielectric constant of material is given by the plasma
energy $\hbar\omega_p$:
\beq
  \epsilon \sim 1 - \left( {\hbar\omega_p\over T}\right)^2\,,
\eeq
where
\beq
    \hbar\omega_p ({\rm eV}) = 28.8\sqrt{{Z\over A}\rho(g/{\rm cm}^3)}\,.
  \label{eq:plasma}
\eeq
For a molecule, $Z$ and $A$ refer to those of the whole molecule; the
plasma energy is proportional to the square root of the electron number
density.
Then, for $\beta\to 1$, we have a plateau at
\beq
   \ln{1\over\sqrt{(1-\beta^2\epsilon_1)^2 + \beta^4\epsilon_2^2}} \to
   \ln{T^2\over (\hbar\omega_p)^2}
\eeq
which the logarithmic rise (\ref{eq:lowden}) would cross at
\beq
   \gamma = {T\over \hbar\omega_p}\,.
  \label{eq:gamsat}
\eeq
This is for a given energy of the exchanged photon.
In practice, the saturation energy is given by replacing 
the recoil energy $T$ by the effective excitation energy $I$:
\beq
   \gamma_{\rm sat} \sim {I\over \hbar\omega_p}\,.
\eeq
For elements, we have approximately~\cite{SeltzerBerger}
\beq
    I ({\rm eV}) \sim 13.5 Z\quad(Z\le 14)\,,\quad
    I ({\rm eV}) \sim 10 Z\quad(Z > 14)\,,
\eeq
which is good to 20\%\ except for $I({\rm H}) = 19.2$ eV 
and $I({\rm He}) = 41.8$ eV.
For compound molecules or gas mixtures,  a good approximation 
is the Bragg's additivity
\beq
     \log I = {\sum_i n_i \log I_i \over \sum_i n_i}\,,
\eeq
where $n_i$ is the total number of electrons 
per molecule of a given element $i$.
For hydrocarbons C$_k$H$_l$, for example, 
$n_C = 6k$ and $n_H = l$, and gives around $45\pm5$ eV
for wide range of values for $k$ and $l$.
For gas mixtures, we have $n_i = Z_i f_i$ where
$Z_i$ is the total number of electrons per $i$-th molecule
and $f_i$ is the fraction by volume. Even though
the effective excitation energy in principle should depend on the type of
chemical bond, the above formula was found to work
for most cases.~\cite{SeltzerBerger}
The heavier the element, and the lower the
density, the higher is the saturation point $\gamma_{\rm sat}$.
Table~\ref{tb:sat} gives the saturation point for some gases as well as
the saturation momentum for $\pi$ ($P_{\rm sat}^\pi$) and $K$ ($P_{\rm sat}^K$).
For a given type of gas,
the $K\pi$ separation starts to degrade as the momentum goes higher than
$P_{\rm sat}^\pi$ and becomes useless around $P_{\rm sat}^K$.
\begin{table}
  \begin{center}
   \begin{tabular}{c|cccc}
    \hline
     \tbst gas & $I$  & $\hbar\omega_p$ & $\gamma_{\rm sat}$  &
      $p_{\rm sat}^{\pi/K}$ \\
     \tbst     &    (eV) &         (eV) &  &  (GeV/c)   \\
    \hline
     \tbst He         & 41.8 & 0.27 & 154  & 21/76  \\
     \tbst Ar         & 188  & 0.82 & 230  & 32/115 \\
     \tbst Xe         & 482  & 1.41 & 341  & 48/170 \\
     \tbst CH$_4$     & 41.7 & 0.61 & 68.4 & 10/34  \\
     \tbst C$_2$H$_6$ & 45.4 & 0.82 & 55.3 & 8/28   \\
     \tbst C$_3$H$_8$ & 47.1 & 0.96 & 49.1 & 7/24   \\
  \tbst C$_4$H$_{10}$ & 48.3 & 1.14 & 42.4 & 6/21   \\
    \hline
   \end{tabular}
   \caption{}
  \end{center}
  \label{tb:sat}
\end{table}

\section{Mean $dE/dx$ vs $\beta\gamma$}

\begin{figure}
 \centering
 \mbox{\psfig{figure=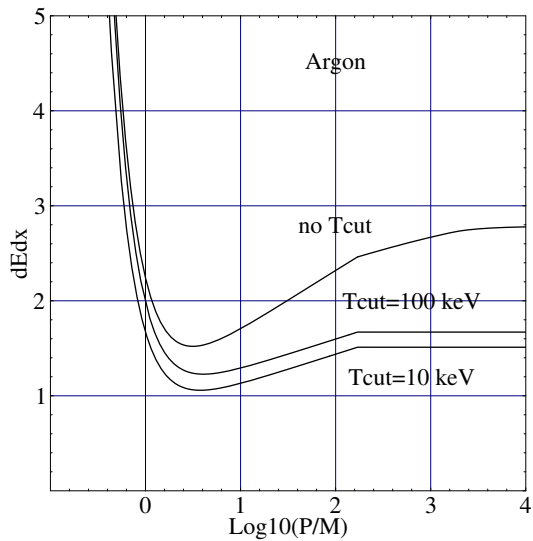,height=2.8in,width=2.8in}}
 \caption{The ionization energy deposit $dE/dx$ (MeV/g cm$^{-2}$)
   for argon plotted as a function of $\beta\gamma$ of projectile.
   Three curves corresponds to no $T_{\rm cut}$, $T_{\rm cut}=100$ keV,
   and $T_{\rm cut} = 10$ keV.}
 \label{fg:dEdx}
\end{figure}

When averaged over the recoil energy,  (\ref{eq:AllisCobb}) eventually
leads to the (truncation-improved) Bethe-Bloch formula:~\cite{PDG}
\beq
   {dE\over dx} \propto {1\over\beta^2}\left(
     {1\over2} \log{2m_e\beta^2\gamma^2 T_0\over I^2} - 
    {\beta^2\over2} \Big( 1 + {T_0\over T_{\rm max}} \Big) - {\delta\over2}\right)
  \label{eq:Bethe}
\eeq
where $T_0 = \min(T_{\rm cut},T_{\rm max})$, $T_{\rm cut}$ is the effective cutoff
for the knock-on electron energy, and $T_{\rm max}$ is its kinematic limit:
\beq
        T_{\rm max} = {2 (\gamma\beta)^2 m_e\over 1 + x^2 + 2 \gamma x}\,,
\eeq
where $x=m_e/M$ with $M$ being the mass of the projectile.
We have $T_{\rm max}\sim P$ for $\gamma x \gg 1$, or the recoil energy can be
as large as the energy of projectile itself.
Such large recoil electrons should clearly not be included in the
evaluation of $dE/dx$ measured along the projectile since they will form
separate tracks. The truncation $T_{\rm cut}$ in the
expression above is designed to take care of such
effect. Typical value for $T_{\rm cut}$ is 10 to 100 keV depending 
on the selection criteria used 
for drift chamber hits. A knock-on electron of 1 MeV will curl up with $r=1.5$ mm
in a 2 Tesla field, and will probably be rejected by tracking code. Also, usually
a fixed fraction of high-side (or sometimes also low-side) tail is rejected in
forming the $dE/dx$ of a track (the truncated mean method) which also leads to
an effective $T_{\rm cut}$.
Figure~\ref{fg:dEdx} plots (\ref{eq:Bethe}) for argon
with no $T_{\rm cut}$, $T_{\rm cut}=100$ keV,
and $T_{\rm cut} = 10$ keV.
The absolute value of the total relativistic rise is
reduced by about factor of two upon applying $T_{\rm cut}$, but is
relatively insensitive to the values of $T_{\rm cut}$ in the range
of 10-100 keV.

\section{Resolution}

Empirically, the resolution of gas-sampling device can be 
approximated by~\cite{Walenta}
\beq
   {\sigma\over\mu} = 0.41 n^{-0.43} (x P)^{-0.32}\,,
\eeq
where $n$ is the number of samples, $x$ the sampling thickness in cm, and
$P$ the pressure in atm. 
This is obtained by a fit to actual resolutions measured, and includes
optimizations of resolution using such techniques as the
truncated mean method.
Another method is to employ a maximum likelihood fit to all the
pulse-heights along a track. Such technique is known to result in similar, if
slightly better, resolutions.
Table~\ref{tb:resol} shows a comparison of some
resolutions measured by recent experiments
and those `expected' from the formula above.
\begin{table}
  \begin{center}
   \begin{tabular}{@{\tbst}c|ccccc}
    \hline
     det. & $n$ & $x$(cm) & $P$ & exp. & meas. \\
    \hline
     Belle    & 52  & 1.5  & 1 atm    & 6.6\%\ & 5.1\%\ ($\mu$) \\
     CLEO2    & 51  & 1.4  & 1 atm    & 6.4\%\ & 5.7\%\ ($\mu$) \\
     Aleph    & 344 & 0.36 & 1 atm    & 4.6\%\ & 4.5\%\ ($e$)   \\
     TPC/PEP  & 180 & 0.5  & 8.5 atm  & 2.8\%\ & 2.5\%\         \\
     OPAL     & 159 & 0.5  & 4 atm    & 3.0\%\ & 3.1\%\ ($\mu$) \\
     MKII/SLC & 72  & 0.833& 1 atm    & 6.9\%\ & 7.0\%\ ($e$)   \\
    \hline
   \end{tabular}
  \caption{Measured resolutions are compared to the 
           expected $dE/dx$ resolutions that are estimated by
           the empirical formula given in the text.}
  \end{center}
   \label{tb:resol}
\end{table}
The formula works well except for CLEO2 and Belle for which the gas contains
a large fraction (50\%) of hydrocarbon (ethane).%
\footnote{A better fit can be obtained by replacing $(xP)$ with
  $6.83 n_e x({\rm cm})  P({\rm atm}) / I ({\rm eV})$ where $n_e$ is the number
  of electron per molecule. See Reference 1.}
Hydrocarbons typically have better resolutions for a given sampling condition $n,x$ and $P$
since they have relatively
small values of mean ionization energy
with respect to the number of electrons per molecule resulting in a larger total
number of ionization.

If each sampling is independent and there are no other sources of error
such as electrical noises, then the resolution would scale as $n^{-0.5}$.
Since the power on $n$ is larger than that on $x$, for a fixed total length
one obtains a better resolution for a finer sampling. At some point, however,
the number of primary ionization which is approximately given by%
\footnote{This formula is usually good to 20\%. Some exceptions are
   5.2 for H$_2$, 5.9 for He, and 44 for Xe.}
\beq
    n_p \sim 1.5 Z/{\rm cm} \quad(Z:\hbox{whole molecule})
\eeq
becomes of order unity and there will be no gain in resolution thereafter.
Typically, the critical sampling size is about a few mm.

\begin{figure}
 \centering
 \mbox{\psfig{figure=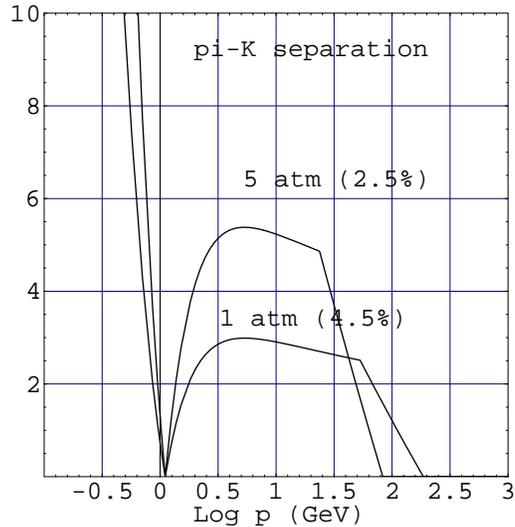,height=2.8in,width=2.8in}}
 \caption{$K/\pi$ separation (number of sigma) for argon as functions of 
   $\log P$(GeV/c). Two curves correspond to 1 atm and 5 atm. }
 \label{fg:piKsep}
\end{figure}

We also notice in Table~\ref{tb:resol} that higher pressure is effective in
improving the resolution. As we have seen in (\ref{eq:gamsat}) and (\ref{eq:plasma}),
however, the saturation energy is proportional to
$1/\sqrt{\rho}$ or equivalently to
$1/\sqrt{P}$ and thus $dE/dx$ saturates at lower energies
for higher pressures. Also, a high-pressure chamber adds problems to
design and fabrication of the tracking device, and also adds more materials
compared to non-pressurized chambers.
To study particle type separation, we take two reference points
for argon: 4.5\%\ resolution at 1 atm and 2.5\%\ at 5 atm. These are quite
realistic values. The resulting $K/\pi$ separation defined as
$(dE/dx|_\pi - dE/dx|_K)/ \sigma_{dE/dx}$ is shown in Figure~\ref{fg:piKsep}
for the two cases.
For 1 atm, 2$\sigma$ separation can be achieved for $p<0.8$ GeV/c and 
$1.75<p<65$ GeV/c. On the other hand for 5 atm, 2$\sigma$ separation can be achieved 
for $p<0.9$ GeV/c and  $1.25<p<50$ GeV/c, and 4$\sigma$ separation can be achieved 
for $p<0.6$ GeV/c and  $1.75<p<30$ GeV/c.

\section{Silicon trackers}

Silicon has $\rho=2.33$ gr/cm$^3$ and $I = 173$ eV. As a result, the saturation
occurs at a low value of $\gamma_{\rm sat} = 5.6$ which can be compared to
the minimum $dE/dx$ location of $\gamma\sim 4$. There is essentially no
logarithmic rise which is a general feature of solid or liquid, and thus
$K/\pi$ separation in the logarithmic rise region is not practical.
It can, however, be used below the minimum ionization region.

Achievable resolution is limited by the number of sampling even though the
total ionization energy in 5 layers of 0.3 mm thick silicon is about the same as
that in 1.5 m of argon gas. 
Since each sampling is `thick', the fluctuation in the Landau tail
degrades the resolution.
A Babar study~\cite{Schumm} based on the Vavilov model of ionization
showed that 10.4 \%\ resolution
can be obtained by 5 layers of silicon strip detectors (0.3 mm thick) for
pions at 450 MeV/c. Also an ALICE study based on 2 layers of silicon strip
detector and 2 layers of silicon drift detector resulted in 10.6 \%\ resolution
for kaons at 0.98 GeV/c. 
In both cases, the numbers correspond to the truncated mean method where
two highest pulse heights were discarded.
Such resolution can give 2$\sigma$ separation for $p<0.65$ GeV/c, and quite
adequate for search of new heavy charged particles such as stau in the gauge-mediated
SUSY breaking models. It should be noted, however, that in order to fully utilize
the low energy region where $dE/dx$ rapidly rise as $1/\beta^2$, the dynamic
range of the readout system should cover up to around 20 times the minimum
value of $dE/dx$.

\section{Summary}

The logarithmic rise of $dE/dx$ and the
saturation energy of the rise are well understood; in
particular, the saturation point is higher for 
heavier element and pressurization results in a lower
saturation. 
The measured $dE/dx$ resolutions can be well approximated by
an empirical formula which can be used as a guide to select the
type of gas. The resolution is usually better for gases with higher
content of hydrocarbons, but they typically have lower saturation
energies. The larger is the number of sampling, the better is the
resolution, but up to the sampling size of a few mm.
For silicon trackers, the logarithmic rise is essentially
absent, and thus only useful region is below the minimum ionizing
point. Still, the achievable resolution is about 10\%\ which is adequate for
detection of new heavy particles such as stau in the gauge-mediated
SUSY breaking models.

\end{document}